# Chapter 8

# Recent advances in superlattice frequency multipliers

**Yuliia Shevchenko, Apostolos Apostolakis, Mauro F. Pereira**

**Abstract** Semiconductor superlattice frequency multipliers have emerged as a nonlinear medium able to generate radiation in a wide frequency range. This property facilitates the potential of sources suitable for sensing and spectroscopy applications. In this study, we further investigate the consequences on harmonic generation in a superlattice multiplier after excitation by an input signal oscillating at different frequencies. Here we provide a rigorous description of our theoretical model including a semiclassical Boltzmann approach to nonlinear miniband transport and non-equilibrium Green's functions calculations treating scattering processes under forward and reverse bias. To fully exploit the features of this radiation source, we focus on the effects of elastic scattering and systematic imperfections in the superlattice structure which lead to asymmetric current flow.

**Keywords**: frequency multiplication, gigahertz, terahertz, superlattices, CBRN agents, harmonic generation, elastic scattering

Yuliaa Schevchenko and Apostolos Apostolakis

Department of Condensed Matter Theory, Institute of Physics CAS, Na Slovance 1999/2 182 21 Prague, Czech Republic
shevchenko@fzu.cz ; apostolakis@fzu.cz

Mauro F. Pereira

Department of Physics, Khalifa University of Science and Technology, Abu Dhabi 127788, UAE and
Department of Condensed Matter Theory, Institute of Physics CAS, Na Slovance 1999/2 182 21 Prague, Czech Republic
pereira@fzu.cz ; mauro.pereira@ku.ac.ae

## 8.1 Introduction

The physical mechanisms of interactions between THz waves and matter can involve lattice vibrations, molecular rotations, spin waves and internal excitations of bound electron-hole pairs [1-4]. It therefore should come as no surprise that the THz technology and applications which might detect the phase and amplitude of generated electromagnetic radiation in the THz range are important components of spectroscopic tools with enhanced temporal and spatial resolution [1, 2]. These properties would be especially desirable for a wide range of chemical, biological, radiological and nuclear defence (CBRN) agents and explosives which exhibit acute sensitivity at GHz, THz and Mid Infrared (TERA-MIR) frequency ranges [5]. This represents further evidence for the growing interest in materials and devices which can detect or generate GHz and TERA-MIR radiation. In particular, semiconductor superlattices (SSLs) [6, 7] are auspicious devices operating at GHz-THz frequencies due to their tuneable electro-optic properties [8-10]. These type of artificial structures are created by periodical layering of two different semiconductor materials with different energy bands which results to the modulation of potential energy only in one direction and the formation of narrow energy minibands. SSL is a strongly nonlinear medium which makes possible to study interesting electrical phenomena such as negative differential conductivity (NDC) and Bloch oscillations which are difficult to be observed in conventional materials [8, 9]. These inherited nonlinear properties are widely regarded as crucial features to develop devices suitable for harmonic generation and amplification of THz radiation. In fact, when Bragg reflected miniband electrons interact with an oscillating electric field, spontaneous frequency multiplication effect takes place. Superlattice frequency multipliers are being developed as a promising approach for the generation of submillimeter to THz waves [9, 12, 13]. A number of recent works have focused on describing THz-GHz nonlinearities covering both “even” and “odd” responses produced from higher-order multipliers based on SSL devices [9, 11, 14, 15]. Among them, we would like to highlight our joint theoretical and experimental study of harmonic generation in a SSL frequency multiplier at room temperature after excitation by a GHz input signal [9]. The theoretical model in this work was formulated by adopting a hybrid approach that combined a nonequibrium Green’s function (NEGF) implementation and using Boltzmann transport equation under relaxation time approximation [9, 11]. This model predicted accurately the unexpected generation of even harmonics for an unbiased SSL. Such kind of response was attributed to imperfections in the structure which led to asymmetric flow and scattering processes under forward and reverse bias. In a follow-up study we investigated the power-conversion efficiency in irradiated SSLs and how the even harmonic output power is affected by controlling the interface quality and thus the interface roughness relaxation processes [14]. Hence, the previous approach was extended by employing the SLs balance equations to incorporate the effect of interface (elastic) scattering and the NEFG input [8, 14, 16, 17]. Further theoretical studies examined the performance of SSL multipliers and their dependence on different input power sources. The latter study confirmed the unique potential of integrating SLED devices with SSL multipliers [18]. A case of particular interest might arise if someone considers an interplay between the asymmetric current flow, applied static voltages and the input oscillating field [19] that might lead to photon-assisted transport [8, 20, 21]. The photon-assisted transport and the absolute negative conductivity (ANC) has been experimentally observed in weakly coupled superlattices in good agreement with theoretical calculations [20, 21].

Before further discussing our specific case of nonlinear behavior, it is worth noting that nonlinear optical effects in semiconductor materials have been strongly investigated in the near infrared and visible spectra, using a combination of semiconductor Bloch equations and NEGF methods for both interband [23-31] and intersubband cases [32-40] with the optical response due to the transitions between well-defined subbands in both quantum wells, superlattices and quantum cascade device. Here in contrast we exploit a different mechanism, in which a current within a single superlattice miniband generates the optical response and nonlinear effects.

The present work further develops the study in Ref. [14] by providing additional information of the output power of the even harmonics generated by a superlattice multiplier resulting from the differences in the interface structure of the superlattice layers. In addition, we revisit the derivation of Balance equations [8, 16, 17] in detail which involves an implicit connection with the NEGF approach. Finally, we study the rectification effects and how they affect the harmonic emission with or without asymmetric current flow. The Chapter is organized as follows. In Sect. 1.2, we discuss the details of the mathematical formalism describing the electron transport and the power output. Section 1.3 reviews our research for the even harmonics generation in quantum superlattices, providing additional data for a straightforward comparison with the power output of the odd harmonics. Section 1.4 provides a short review of photon-induced transport in SSLs and a preliminary study for the related corrections due to rectification effects. Concluding remarks are given in the end.

## 8.2 Mathematical Formalism

We focus on the nonlinear current response in the case of the tight-binding dispersion relation $\varepsilon_q = \varepsilon(q) = E_a - 2\,|T_1^\alpha|\cos q\,d$ where $q$ is the crystal momentum, $d$ is the lattice period of the SSL and $E_a$ designates the center of the first miniband [8, 9]. The dispersion relation $\varepsilon_q$ can be obtained for typical SSL parameters and next-neighbor coupling $T_1^\alpha$ [8].

Considering a superlattice which is excited by an oscillating field $E(t) = E_{dc} + E_{ac}\cos(2\pi\nu t)$ which is parallel to its growth direction, one can get the general current response and its spectral components [8, 9]

$$j(t) = j_{dc} + \sum_{l=-\infty}^{\infty} j_l^c \cos(2\pi\nu l t) + j_l^s \sin(2\pi\nu l t)\ ,$$

$$j_{dc} = \sum_{p=-\infty}^{\infty} J_p^2\,(\alpha) Y(U),$$

$$j_l^c = \sum_{p=-\infty}^{\infty} J_p(\alpha)[J_{p+l}(a) + J_{p-l}(a)]\ Y(U), \tag{8.1}$$

$$j_l^s = \sum_{p=-\infty}^{\infty} J_p(\alpha)[J_{p+l}(a) - J_{p-l}(a)]\ K(U).$$

Here $J_p$ represents the Bessel function of the first kind and order $p$. The parameter $\alpha = eE_{ac}\,d/\hbar\omega$ designates the modulation degree of Bloch frequency oscillations and therefore controls effectively the nonlinear response of the system [12]. This scheme holds for miniband transport within relaxation time approximation and functions $Y$, $K$ given by

$$Y(U) = j_0\,\frac{2U/\Gamma}{1+(U/\Gamma)^2},\ K(U) = \frac{2j_0}{1+(U/\Gamma)^2}, \tag{8.2}$$

where $U = eE_{dc}d + p\hbar\omega$ is the total energy drop per period under irradiation. The parameter $j_0$ denotes the peak current whereas $\Gamma = \hbar/\tau$ is the scattering induced broadening. Both of these parameters can be microscopically calculated by implementation of NEGF approach, as in Refs. [9, 11]. It is crucial to note that it is possible to consider the domain-formation effects in superlattices under irradiation [8, 13]. This situation is typically unstable with respect to the formation of inhomogeneous field distributions and starts to appear if $a > a_c = U_c/h\nu$. The latter condition corresponds to the NDC state and the onset of Bloch oscillations. However, in this work we restrict ourselves by studying the current response in the absence of electric field domains. It is well-established that in absence of domains, or of any other asymmetry, the *I-V* characteristic can be described adequately by Eq. (8.2) with considering a purely dc biased SL. One can identify two basic aspects of miniband transport under irradiation by means of harmonic radiation:

(i) For a perfectly symmetric structure only odd harmonics can appear if $E_{dc} = 0$.

(ii) For a superlattice structure with asymmetric interfaces both even and odd harmonic orders can occur, indicating that spectral tuning of high-order harmonics is feasible beyond the parametric amplification mechanisms [40].

These effects were revealed by NEGF approach by including systematically the possibility that interface of GaAs over AlAs is worse than that of AlAs over GaAs [9]. This resulted in a deviation from the standard Esaki-Tsu model which can be represented by the following ansatz

$$j_0 = \begin{cases} j_0^-, & U < 0 \\ j_0^+, & U \geq 0 \end{cases}, \qquad \Gamma = \begin{cases} \Gamma^-, & U < 0 \\ \Gamma^+, & U \geq 0. \end{cases} \tag{8.3}$$

The above compact solution implies that the parameters $j_0^+$, $j_0^-$, i.e. the maximum and minimum current density, correspond to the critical energies $U_c^+$ and $U_c^-$. A direct connection between calculated global dephasing and these extrema is given by $\Gamma^+ = U_c^+$ and $\Gamma^- = U_c^-$.

To determine the power output of a superlattice multiplier in the presence of an ac field, we calculate the average emitted power due to the harmonic which is directly obtained from the Poynting vector [11, 14]. Hence, the generated power can be expressed as

$$P_l\,(\omega) = \frac{A\,\mu_0\,c\,L^2}{8\,\mathrm{n_r}} \mathcal{J}_l^2(\nu), \tag{8.4}$$

where $A$ is the area of the mesa of a superlattice element, $\mu_0$ is the permeability of the vacuum and $c$ is speed of light in the vacuum, $L$ is the effective path length through the crystal and $n_r$ is the refractive index of the SL material. The term $\mathcal{J}_l(\nu)$ is the root-mean-square value of the $l^{\text{th}}$ component of the expansion of the induced current density in Eq. (8.1).

### 8.2.1 Balance equations and effect of interface roughness scattering

The dynamical evolution of distribution function $\mathrm{f}(q, \mathbf{k}, t)$, which can be analyzed by adopting the semiclassical theoretical framework , is determined by Boltzmann transport equation (BTE) [8]

$$\frac{\partial \mathrm{f}(q,\mathbf{k},t)}{\partial t} + \frac{\mathrm{e}E(t)}{\hbar}\frac{\partial \mathrm{f}(q,\mathbf{k},t)}{\partial \mathrm{q}} = St(\mathrm{f}), \tag{8.5}$$

where $E(t)$ is the input oscillating field. To overcome the limitations of the relaxation-time model we employ a collision integral which can describe additionally the elastic lattice scattering and leads to the following formal definition [8, 16, 17]

$$St(\mathrm{f}) = -\frac{f(q,\mathbf{k},t) - n_{\mathrm{F}}(\mathrm{E}(q,\mathbf{k}) - \mu)}{\tau_\varepsilon} + \frac{f(-q,\mathbf{k},t) - f(q,\mathbf{k},t)}{2\tau_{int}}\,. \tag{8.6}$$

Here, we consider the effects of different scattering processes which affect the electron energy and the miniband electron velocity; they are contained within the phenomenological scattering constants $\tau_\varepsilon$ and $1/\tau_v = 1/\tau_\varepsilon + 1/\tau_{int}$, where $\tau_{int}$ is the scattering rate related to the elastic processes, namely the interface scattering with which we are primarily concerned in the current study. The distribution function in thermal equilibrium and in the absence of external fields is described by Fermi-Dirac distribution $n_{\mathrm{F}}\,(E(q,\boldsymbol{k})-\mu)=\frac{1}{e^{\frac{[\varepsilon(q)-\mu]}{k_B T_e}}+1}$. Now, the dynamical equations of the average miniband energy and the current density are determined by the distribution function $f(q,\mathbf{k},t)$ as

$$w(t)=\frac{2}{(2\pi)^3 n_e}\iint_{-\infty}^{\infty} d\boldsymbol{k}\int_{-\frac{\pi}{d}}^{\frac{\pi}{d}} dq\;\varepsilon_q\, f(q,\mathbf{k},t), \tag{8.7a}$$

$$J(t)=\frac{2e}{(2\pi)^3}\iint_{-\infty}^{\infty} d\boldsymbol{k}\int_{-\frac{\pi}{d}}^{\frac{\pi}{d}} dq\; v(q)\, f(q,\mathbf{k},t), \tag{8.7b}$$

where $n_e(t)$ denotes the density of electrons and again is determined by the distribution function

$$n_e(t)=\frac{2}{(2\pi)^3}\iint_{-\infty}^{\infty} d\boldsymbol{k}\int_{-\frac{\pi}{d}}^{\frac{\pi}{d}} dq\; f(q,\mathbf{k},t). \tag{8.8}$$

To obtain the dynamical equations of the above-mentioned physical quantities, we start by multiplying Eq. (8.5) by either $\varepsilon_q$ or $v(\mathrm{q})$ and then performing the integral $\frac{2}{(2\pi)^3 n_e}\iint_{-\infty}^{\infty} d\boldsymbol{k}\int_{-\frac{\pi}{d}}^{\frac{\pi}{d}} dq$ and $\frac{2e}{(2\pi)^3}\iint_{-\infty}^{\infty} d\boldsymbol{k}\int_{-\frac{\pi}{d}}^{\frac{\pi}{d}} dq$ , respectively, over the same equation. In particular, multiplying by miniband energy gives the following equation

$$\frac{\partial}{\partial \mathrm{t}}\frac{2}{(2\pi)^3 n_e}\iint_{-\infty}^{\infty} d\boldsymbol{k}\int_{-\frac{\pi}{d}}^{\frac{\pi}{d}} dq\;\varepsilon_q f(q,\mathbf{k},t)+\frac{\mathrm{e}E(t)}{\hbar}\frac{2}{(2\pi)^3 n_e}\iint_{-\infty}^{\infty} d\boldsymbol{k}\int_{-\frac{\pi}{d}}^{\frac{\pi}{d}}\frac{\partial f(q,\mathbf{k},t)}{\partial \mathrm{q}}\varepsilon_q dq\;=$$
$$\frac{1}{\tau_\varepsilon}\frac{2}{(2\pi)^3 n_e}\iint_{-\infty}^{\infty} d\boldsymbol{k}\int_{-\frac{\pi}{d}}^{\frac{\pi}{d}} n_{\mathrm{F}}\;\varepsilon_q dq-\frac{1}{\tau_\varepsilon}\frac{2}{(2\pi)^3 n_e}\iint_{-\infty}^{\infty} d\boldsymbol{k}\int_{-\frac{\pi}{d}}^{\frac{\pi}{d}} dq\;\varepsilon_q f(q,\mathbf{k},t)-$$
$$\frac{1}{2\tau_{int}}\frac{2}{(2\pi)^3 n_e}\iint_{-\infty}^{\infty} d\boldsymbol{k}\int_{-\frac{\pi}{d}}^{\frac{\pi}{d}} dq\;\varepsilon_q[f(-q,\mathbf{k},t)-f(q,\mathbf{k},t)] \tag{8.9}$$

The third term on the right hand side of the Eq. (8.9) would be identically zero since the product $\varepsilon_q[f(-q,\mathbf{k},t)-f(q,\mathbf{k},t)]$ is an odd function. Using the equations (8.7a), (8.7b), (8.9) and the integration by parts for the second term on the left hand side of equation (8.9), we finally obtain

$$\dot{w}(t)-\frac{E(t)J(t)}{n_e}=\frac{\mathrm{w}_{\mathrm{eq}}(\mu,T)-w(t)}{\tau_\varepsilon} \tag{8.10}$$

with the miniband energy in thermal equilibrium in the absence of an external electric field given by

$$w_{\mathrm{eq}}(\mu,T)=E_a-\frac{2}{(2\pi)^3 n_0}\iint_{-\infty}^{\infty} d\boldsymbol{k}\int_{-\frac{\pi}{d}}^{\frac{\pi}{d}} dq\; 2\,T\cos(qd)n_{\mathrm{F}}. \tag{8.11}$$

In this case the electron density is obtained from Eq. (8.8), which can be represented with

$$n_e=n_0=\frac{2}{(2\pi)^3}\iint_{-\infty}^{\infty} d\boldsymbol{k}\int_{-\frac{\pi}{d}}^{\frac{\pi}{d}} dq\; f(q,\mathbf{k},t).$$

Similarly, multiplying Eq. (8.5) by $v(\mathrm{q})$ we obtain

$$\frac{\partial}{\partial t}\frac{2e}{(2\pi)^3}\iint_{-\infty}^{\infty} d\boldsymbol{k}\int_{-\frac{\pi}{d}}^{\frac{\pi}{d}} dq\, v(q)\, f(q,\mathbf{k},t) + \frac{eE(t)}{\hbar}\frac{2e}{(2\pi)^3}\iint_{-\infty}^{\infty} d\boldsymbol{k}\int_{-\frac{\pi}{d}}^{\frac{\pi}{d}} \frac{\partial f(q,\mathbf{k},t)}{\partial q} v(q)dq \;=$$
$$\frac{1}{\tau_\varepsilon}\frac{2e}{(2\pi)^3}\iint_{-\infty}^{\infty} d\boldsymbol{k}\int_{-\frac{\pi}{d}}^{\frac{\pi}{d}} n_{\mathrm{F}}\, v(q)dq - \frac{1}{\tau_\varepsilon}\frac{2e}{(2\pi)^3}\iint_{-\infty}^{\infty} d\boldsymbol{k}\int_{-\frac{\pi}{d}}^{\frac{\pi}{d}} dq\, v(q) f(q,\mathbf{k},t) -$$
$$\frac{1}{2\tau_{int}}\frac{2e}{(2\pi)^3}\iint_{-\infty}^{\infty} d\boldsymbol{k}\int_{-\frac{\pi}{d}}^{\frac{\pi}{d}} dq\, v(q)[f(-q,\mathbf{k},t) - f(q,\mathbf{k},t)] \tag{8.12}$$

Here again

$$\dot{J}(t) + \frac{d^2 eE(t)}{\hbar^2}[w(t) - E_a] = -\frac{J(t)}{\tau_v} \tag{8.13}$$

The balance equations (8.10), (8.13) describe the dynamics of a quantum particle in a miniband of a semiconductor superlattice [16, 17]. The Bragg reflections are consistently included in the description of these dynamical equations by assuming the condition $f(-\pi/d,\mathbf{k},t) = f(\pi/d,\mathbf{k},t)$ for the distribution function. Hereafter, we construct the stationary solution of the dynamical equations (8.10), (8.13) by assuming that a static field $E_{dc}$ is applied and therefore we obtain

$$J(U) = \beta\, j_0 \frac{2\frac{U\tau_{eff}}{\hbar}}{1+\left(\frac{U\tau_{eff}}{\hbar}\right)^2} \tag{8.14}$$

with the effective scattering time $\tau_{eff} = \sqrt{\tau_\varepsilon \tau_v}$ and $\beta = \sqrt{\tau_v/\tau_\varepsilon}$ and the magnitude of the peak current $j_0 = (edn_0/2\hbar)(E_a - w_{eq})$ . The connection between voltage and energy is $U = eV = eE_{dc}d$. In a similar way we obtain the relationship that describes how far from the middle of the miniband is the average energy of electrons

$$\frac{\Delta W(U)}{\Delta W_0} = \frac{1}{1+\left(\frac{\hbar}{U\tau_{eff}}\right)^2}. \tag{8.15}$$

Here $\Delta W = (w - w_{eq})$ and $\Delta W_0 = (E_\alpha - w_{eq})$.

The voltage-current characteristic corresponding to the analytical formula of Esaki-Tsu curve is illustrated in figure 8.1(a). The magnitude of the peak current is decreased by the factor $\beta$ in comparison with Eq. (8.1). By increasing the voltage beyond the point where the current obtains its maximum value, the electrons are allowed to proceed along the dispersion curve $\varepsilon_q(q)$, and finally to reach the edge of Brillouin zone. This results in the suppression of electron transport which is manifested in the reduction of current density with the increase of $U$. On the contrary, the average minband energy increases with increasing $U$. In particular, in the limit $U \gg U_{cr}$ the fraction $\Delta W/\Delta W_0$ approaches unity and therefore $W$ is located in the vicinity of the center of the miniband ( $W \rightarrow E_a$ ). An important practical question arises: how the balance equations approach is naturally linked with the hybrid approach we discussed earlier in the previous section. The approach based on the SSL balance equations is already known to offer equivalent solutions with the Boltzmann equation in the absence of magnetic field [8]. Thus, we now turn to the consideration of the effects which are related to the asymmetric scattering processes under forward and reverse bias. The presence of the ansatz should naturally lead to a new approximation of the collision integral into the following form

$$St^{as}(\mathrm{f}) = -\frac{f(q,\mathbf{k},t) - n_{\mathrm{F}}(\mathrm{E}(q,\mathbf{k}) - \mu)}{\tau_\varepsilon} + \frac{f(-q,\mathbf{k},t) - f(q,\mathbf{k},t)}{2\tau_{int}^{as}} \tag{8.16}$$

with

$$\tau_{int}^{a_s} = \frac{\tau_{int}}{a_s\,\theta(\mp sgn(v))+1} \qquad (8.17)$$

Such kind of elastic scattering allows to connect in a straightforward manner the effective scattering time $\tau_{eff}$ with the parameter Γ predicted by the NEGF calculations and resulted in the ansatz Eq. (8.3) [14]. The parameter $a_s$ is related to the parameters used to describe the characteristics of each interface and the interface roughness self-energy. The negative (positive) sign in theta function [see Eq. (8.17)] is directly determined by the polarity of the electric field which allows a larger peak current $j_0^+(j_0^-)$.

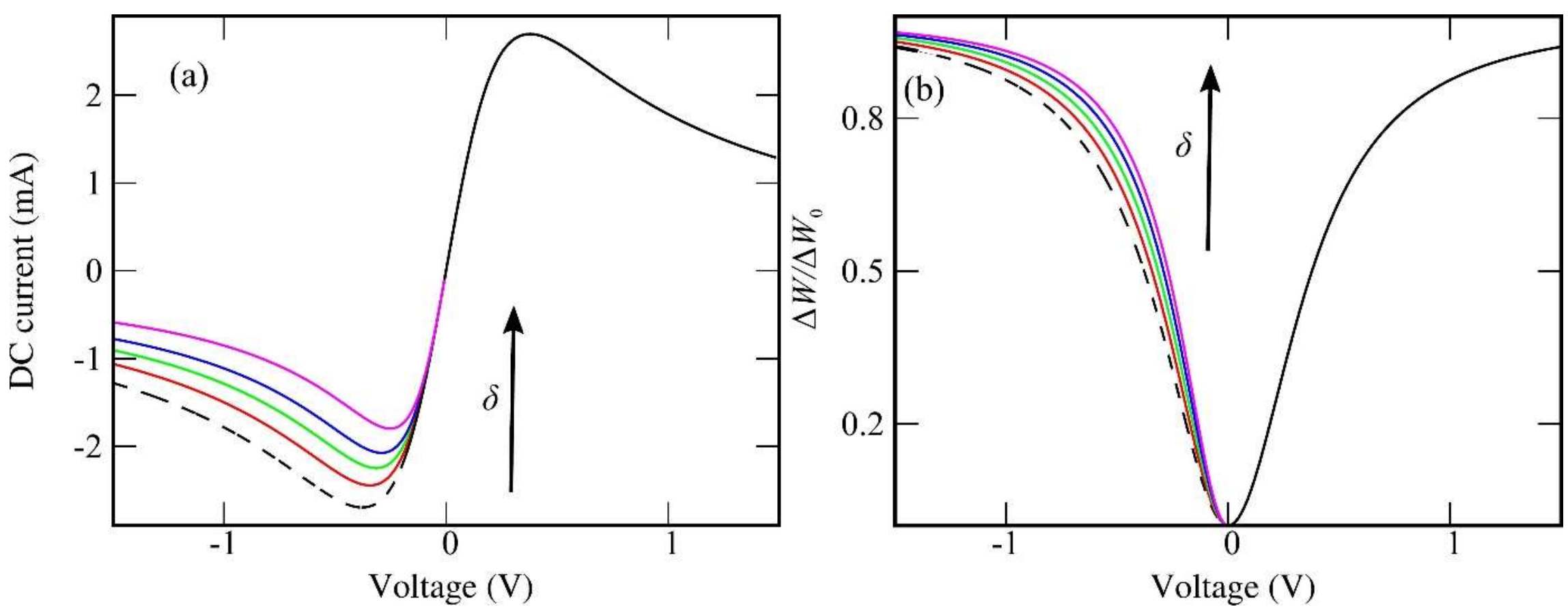

**Fig. 8.1** (a) Current-voltages curves calculated with a variation of the asymmetry parameter which increases by $\delta = 1, 1.1, 1.2, 1.3,\ 1.4$ (b) The average miniband energy $W$ as a function of the applied voltage. The curves above the dashed line (symmetric superlattice) calculated for different asymmetry parameter same as *I-V* curves in Fig. 8.1(a).

Then, the existence of an asymmetric interface roughness design dictates necessary corrections directly in balance equations (8.10), (8.13) which are incorporated as follows

$$j_0 = \begin{cases} \beta^+ j_0, & U < 0 \\ \beta^- j_0, & U \geq 0 \end{cases}$$

where the factor $\beta^+(\beta^-)$ simply reduces (increases) the maximum (minimum) current $j_0$. Therefore, the resulting ansatz permits the use of a parameter $\delta = \beta^-/\beta^+ = (\tau_{eff}^-/\tau_{eff}^+)$ which controls the asymmetry in the current flow and can be obtained from

$$\frac{j_0^+}{\Gamma^+} = \frac{j_0^-}{\Gamma^-}. \qquad (8.18)$$

Here $\delta = j_0^+/\,j_0^-$ and for a perfectly symmetric superlattice $(\delta = 1)$ the corresponding relaxation time is $\tau =$ 31 fs. This way the differences in the SSL interfaces can lead to deviation from a completely antisymmetric current-voltage characteristic. The parameters extracted from NEGF calculations (see Refs. [9, 11]) and used in Eqs. (8.1), (8.2) are: $\Gamma^+$, $\Gamma^-$ =21, 20 meV and $j_0^+$, $j_0^-$=2.14, 1.94× $10^9$ A/m$^2$.
The peak current $j_0^-$ decrease significantly with increasing $\delta$ and therefore results in a strong enhancement of the asymmetry observed in current-voltage curves. This kind of situation is illustrated in Fig. 8(a). Thus, applying static voltage may lead to increase of average miniband energy depending on the field direction as shown in Fig.

8.1(b). These effects which appear due to processes of intraminiband electron scattering can most likely be utilized in generation and amplification of radiation of high-frequency radiation in SSLs [8]. In this work, to estimate the generated power efficiency we will use the following parameters for a GaAs/AlGaAs SSL: period $d$=6.23 nm, electron density $n_0 = 1.5 \times 10^{18}$ cm$^{-3}$ and refractive index $n_r = \sqrt{13}$ (GaAs).

## 8.2 Even-harmonic generation due to designed structural variations of SSLs.

In this section, we analyze the superlattice response to a monochromatic field and consider the effects in terahertz radiation by varying the parameter $\delta$ within reasonable limits. We previously modelled the microscopic frequency multiplication mechanism by determining the conversion efficiency for both even and odd order nonlinearities [14]. However, these calculations did not address a direct comparison between the power output of even harmonics and odd harmonics. To answer this question, we study the response to an ac electric field oscillating at different frequencies and for different asymmetry coefficients $\delta$. Figure 8.1 demonstrates the output powers of the even-order harmonics (second and fourth harmonics) as a function of parameter $\alpha$ in response to an input field which oscillates at 150 GHz. These frequency-doubling and frequency-quadrupling effects would not appear in a perfectly symmetric SSL. On the other hand, as the asymmetry increases the generated harmonics become comparable with the third harmonics produced by a superlattice subjected to irradiation with input fields at different frequencies [see Fig. 8.1(a) and Fig. 8.1(b)]. But, of course, when electrons are able to be efficiently Bragg-reflected, the power output of the third harmonics surpass clearly the even-harmonic radiation.
Note that the NDC state in SSLs is accompanied by an instability of homogeneous electric field resulting in formation of high-field domains. These electric domains can affect dramatically the spontaneous frequency multiplication. Figure 8.2 demonstrates the power emitted by second harmonics, but now after the excitation of the SL sample by higher frequency input signals. Here we notice that the enhanced asymmetry increases considerably the component of the second harmonic generation for frequencies at 300 GHz and 600 GHz. In this case, the excitation of a perfectly symmetric superlattice with an input signal close to 100 GHz should generate a weaker power output sensitive to a wide range of $\alpha$ parameter values, even beyond $\alpha_c$. All the results presented in this section were obtained to complement the effects predicted in [9, 11, 14, 15].

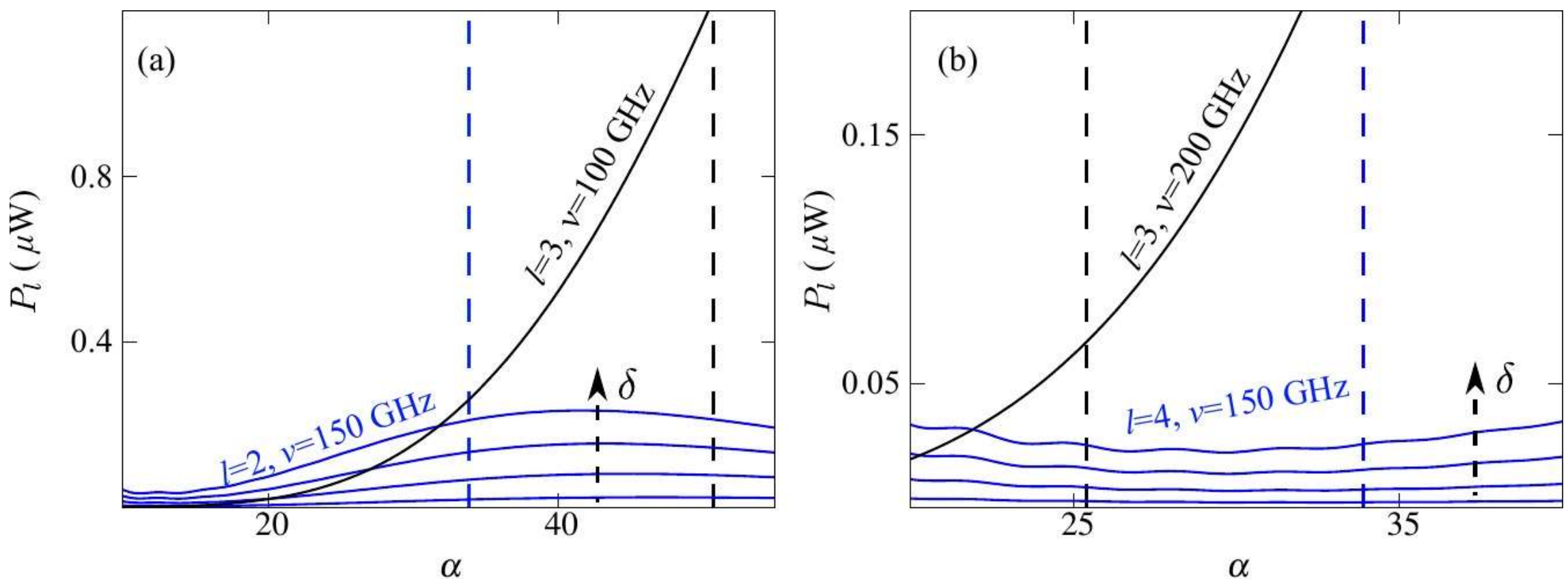

**Fig. 8.2** Comparison of the power output generated by frequency multipliers consisted of either symmetric superlattice or asymmetric ones. (a) Emitted optical power for third harmonic generated by a field oscillating 100 GHz and second harmonic generated by a field oscillating 150 GHz. (b) Emitted optical power for third harmonic generated by a field oscillating 200 GHz and fourth harmonic generated by a field oscillating 150 GHz. The blue curves have been calculated using our theory and varying the asymmetry parameter $\delta = \Gamma^{+}/\Gamma^{-}$ , which increases by $\delta$ =1.1, 1.2, 1.3, 1.4. Note that parameter $\alpha$ determines the input power inside the SSL.

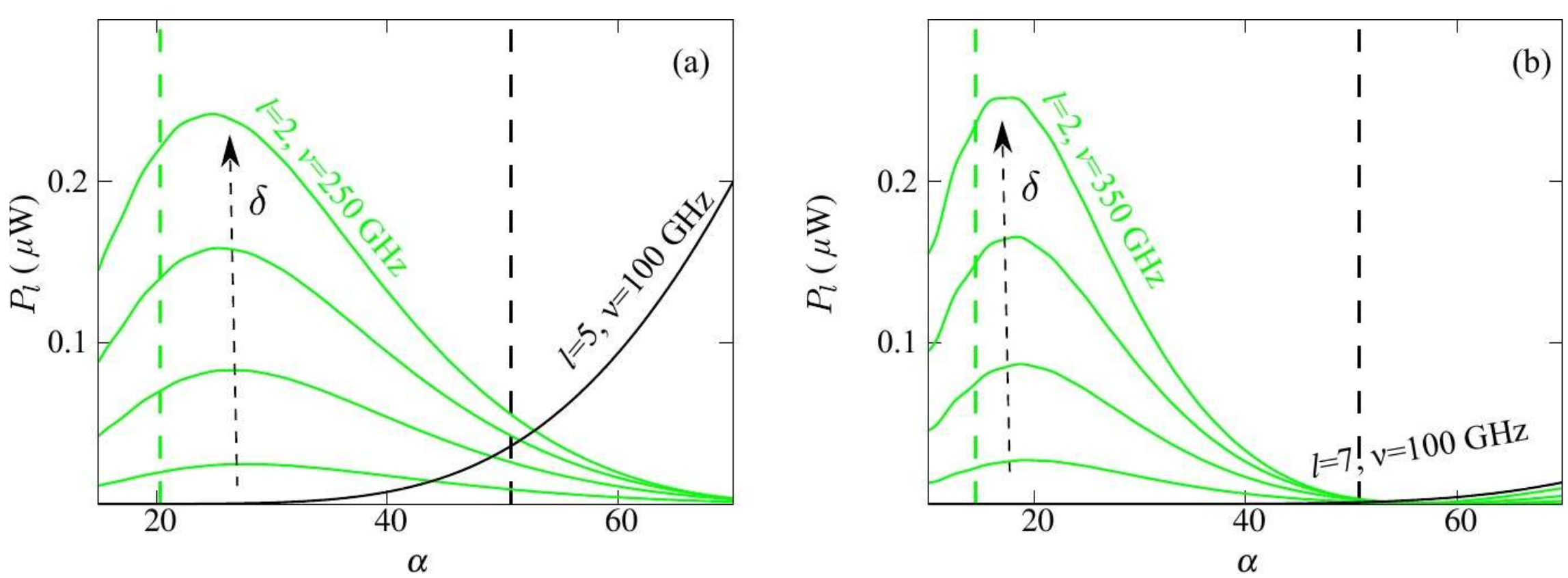

**Fig. 8.3** Comparison of the power output generated by frequency multipliers consisted of either symmetric superlattice or asymmetric ones. (a) Emitted optical power for fifth harmonic generated by a field oscillating 100 GHz and second harmonic generated by a field oscillating 250 GHz. (b) Emitted optical power for seventh harmonic generated by a field oscillating 100 GHz and second harmonic generated by a field oscillating 350 GHz. The blue curves have been calculated using our theory and varying the asymmetry parameter $\delta = \Gamma^{+}/\Gamma^{-}$ , which increases by $\delta$ =1.1, 1.2, 1.3, 1.4. Note that parameter $\alpha$ determines the input power inside the SSL.

## 8.3 Photon-assisted transport and harmonic emission.

Here we discuss the dc-current density in the presence of the ac field, the possibility of formation of photon-assisted peaks and the effects of current rectification on frequency multiplication. The principle underlying SSL multiplication and rectification [8] can be summarized as follows: perpendicular charge transport in biased superlattices is dominated by resonances due to the alignment of energy levels in different wells. These resonances yield various peaks in the current-voltage (I -V) characteristics connected with negative differential conductivity [8, 20]. If the SSL is subjected to an external oscillating field, photon assisted tunnelling is possible and replica of the resonances are observed at biases which differ from alignment conditions by integer multiples of the photon energy. Such kind of resonances are demonstrated in Fig 8.4(a) at $eE_{dc}d = ph\nu + \Gamma$ where $p$ is positive integer. The current voltage characteristics produced by the model calculations with 10.6 THz radiation. Note that these resonance peaks have magnitude proportional to $J_p^2(\alpha)$. There is however another effect which arises, namely the ANC, when the direction of the dc current is opposite to its direction in a normal conductor [e.g. yellow curve in Fig. 8.4(a)]. Furthermore, the physical origin of ANC and the strong suppression of the dc current [red curve in Fig. 8.4(a)] at small values of static bias can be attributed to dynamical localization, when the electron periodically returns to the initial state both in real and quasi-momentum space [21,22]. The condition for dynamical localization is $J_0(\alpha) = 0$ [22]. It should be noticed the maximum response of harmonics corresponds to the resonant condition where both harmonic motions come to resonance, i.e. the one corresponding to the Bloch oscillations due to static voltage $\nu_B = eE_{dc}d/h$ and other one due to frequency modulated Bloch oscillations. Figures 8.4(b)-(d) demonstrate the effectiveness of this condition, i.e. $\nu_B = l\nu/2$ where $l$ denotes the order of the harmonic as we have already discussed. However, we underline that dynamic localization, ANC and the synchronization of the ac field with Bloch oscillations is valid in the limit $\frac{h\nu}{\Gamma} > 1$. Now we further examine the effect asymmetric current flow on the rectification and eventually on the harmonic output. Figure 8.5(a) shows that behavior of the I-V characteristic only slightly changes with the increase of the asymmetry parameter and the emergence of ANC is hardly affected. However, for $E_{dc} \to 0$ the current is clearly rectified and it demonstrates small but non-zero values. As a result, the peak power output for the 4$^{th}$ harmonic notably increases whereas the response of the 2$^{nd}$ and 3$^{rd}$ harmonics remain almost the same [see Figs. 8.4(b)-(d)].

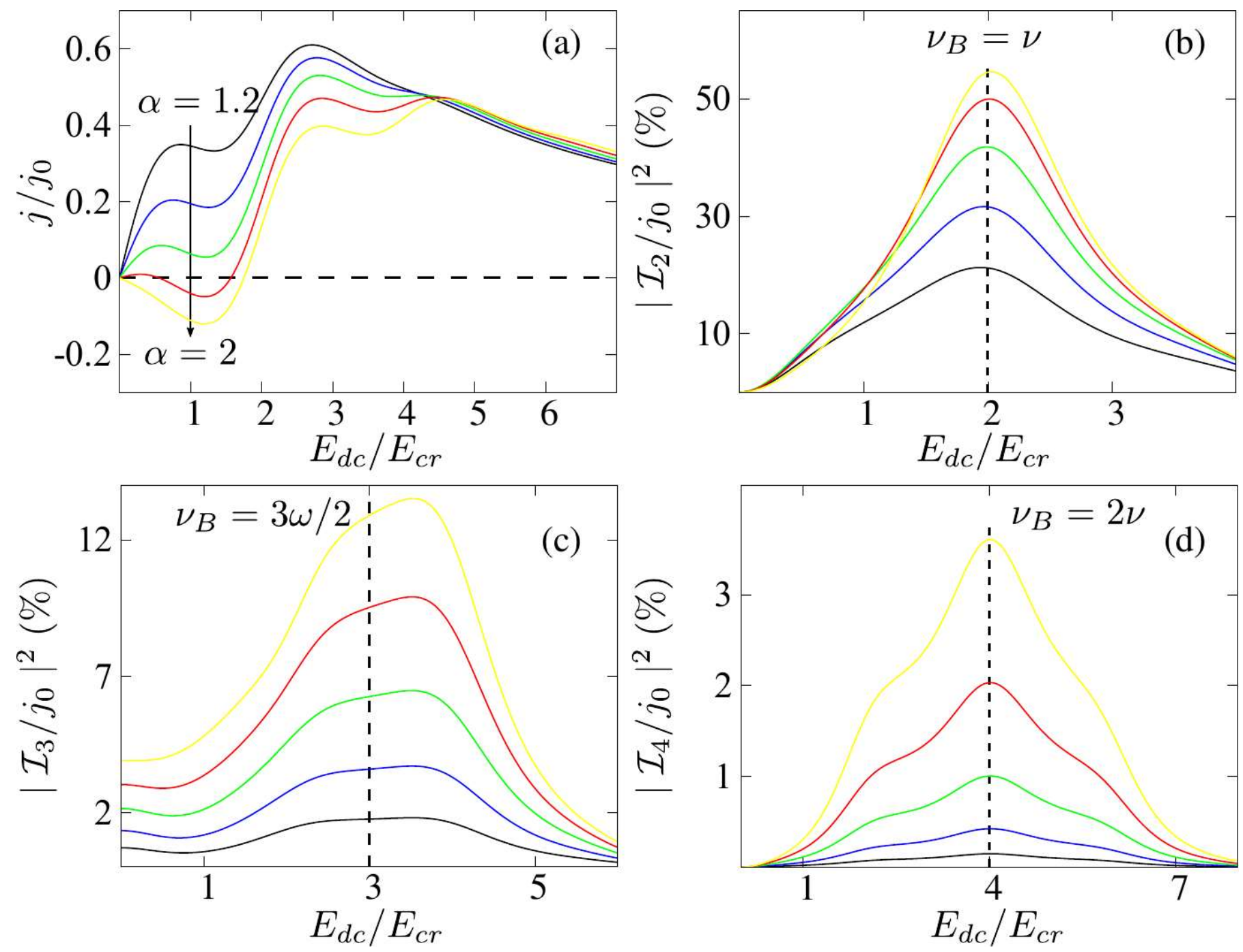

Fig. 8.4 (a) I-V characteristic calculated with input radiation at oscillating frequency $\frac{h\nu}{\Gamma} = 2$. The different colors designate different amplitude of the parameter $\alpha = 1.2, 1.4, 1.6, 1.8, 2$. (b)-(d) The dependence of the harmonic emission $|\mathcal{J}_l(\nu)|^2$ on the static electric field and the different values of parameter $\alpha = eE_{ac}d/(h\nu)$ which causes photon-assisted transport as shown in the panel (a). From (b)-(d) the harmonics are, respectively the 2nd, 3rd and 4th. The vertical dashed lines designate the maximum harmonic output at $\nu_B = l\nu/2$.

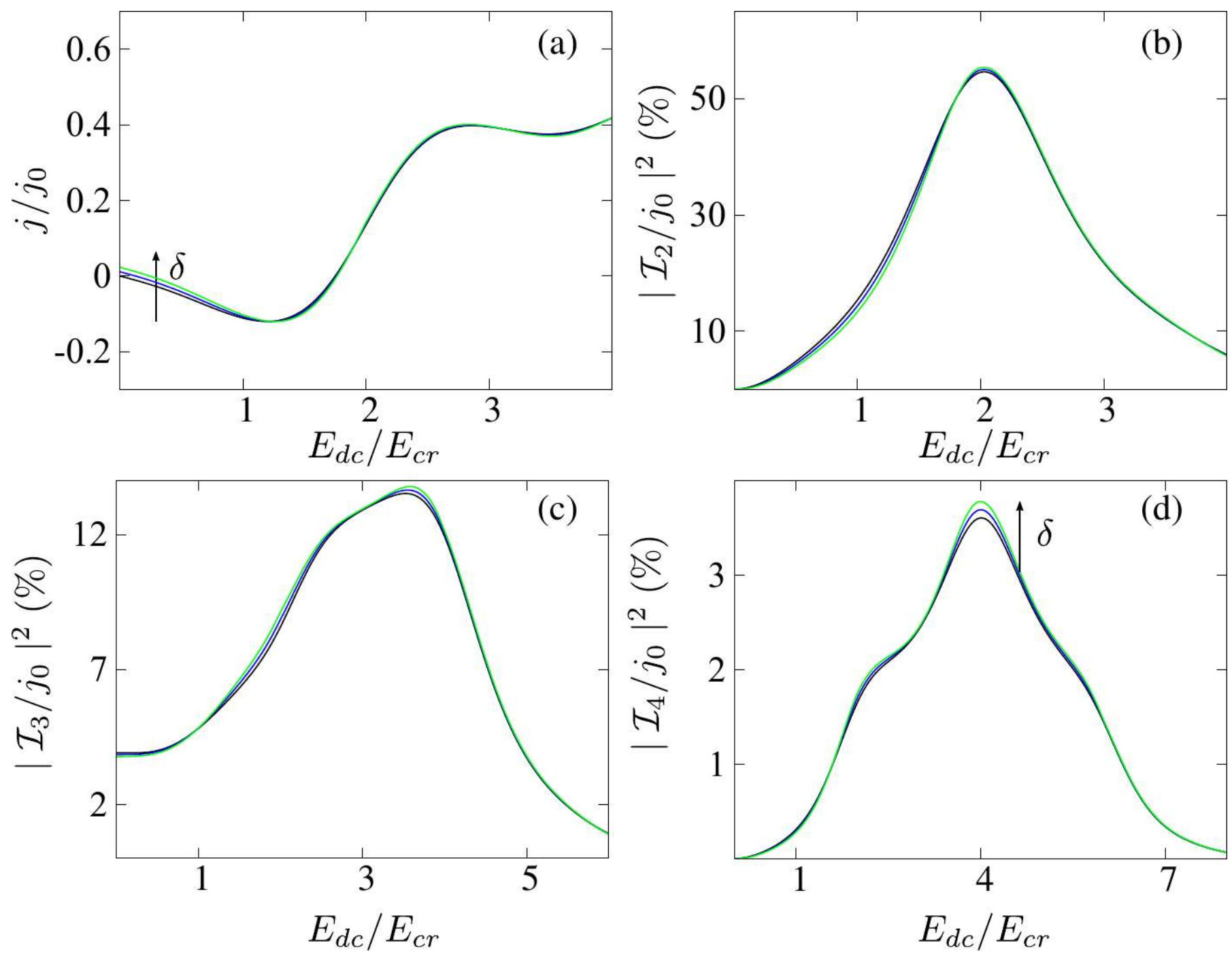

Fig. 8.5(a) I-V characteristic calculated with input radiation at oscillating frequency ($\frac{h\nu}{\Gamma} = 2, \nu = 10.6$ Thz) $\alpha = 2$ and different values of the asymmetry parameter $\delta$. (b)-(d) panels show the dependence of the harmonic emission $|\mathcal{J}_l(\nu)|^2$ on the static electric bias $E_{dc}$. From (b)-(d) the harmonics are, respectively the 2nd, 3rd and 4th. From bottom to the top asymmetry parameter increases by $\delta = 1.05, 1.1$. The black lines correspond to calculations for an ideal superlattice ($\delta = \frac{\Gamma^+}{\Gamma^-} = 1$).

Before stating our concluding remarks, it is noteworthy to consider that SSLMs, given the recent theoretical and experimental advances [42-43], can be used as sources for imaging on medical and security applications, but that the resolution is in principle not as sharp due to the larger wavelengths in the GHz-THz range. However, this may be an opportunity to combined designs with superlensing structures, which can dramatically improve both the transmission and the resolution of the imaged signal [44-46].

## 8.3 Conclusion

In this chapter we revisited the recent progress that has occurred in controllable nonlinear optics for superlattice multipliers in the GHz-THz ranges. Using an extended Boltzmann kinetic approach, we obtained an asymmetry ratio (16) which is consistent with the solution stemmed from NEGF calculations. This study further confirms the potential advantages of generating THz radiation with controlling the superlattice interfaces quality and numerically estimating the nonlinear electromagnetic response. As a further development of this activity, we also presented here preliminary results investigating the rectification effects combined with asymmetrical current flow on the harmonic response for far-infrared wavelength radiation. Currently we are performing different numerical simulations in order to include self-consistently the effect domains in a semiconductor superlattice under the action of GHz field. A different output power can be expected at various frequency ranges by integrating a superlattice multiplier with a SLED oscillator [10]. In order to obtain a more accurate estimation of the conversion efficiency for an irradiated SSL, the effects of different designs of external waveguide delivering the input field power should be considered.